\documentclass[twocolumn,epjc3]{svjour3}
\usepackage{amsmath}
\usepackage{cite}
\usepackage{graphicx,wrapfig,float,slashed}
\usepackage{amsmath,amssymb,epsfig,graphicx}
\usepackage{multirow}
\usepackage{ragged2e}
\usepackage{array}
\newcommand {\tanb}{\tan\beta}

\smartqed

\journalname{Eur. Phys. J. C}

\begin{document}

\title{Analysis of top quark pair production signal from neutral 2HDM Higgs bosons at LHC}

\author{Majid Hashemi \thanksref{e1,a}
\and
Mahbobeh Jafarpour \thanksref{e2,a}
}

\thankstext{e1}{e-mail: majid.hashemi@cern.ch}
\thankstext{e2}{e-mail: mahboob$\_$72$\_$00@yahoo.com}

\institute{Physics Department, College of Sciences,
Shiraz University, Shiraz, 71946-84795, Iran \label{a}
}

\maketitle

\begin{abstract}
In this paper, the top quark pair production events are anaylzed as a source of neutral Higgs bosons of two Higgs doublet model type I at LHC. The production mechanism is $pp \rightarrow H/A \rightarrow t\bar{t}$ assuming a fully hadronic final state through $t\rightarrow Wb \rightarrow jjb$. In order to distinguish the signal from the main background which is the standard model $t\bar{t}$, we benefit from the fact that the top quarks in signal events acquire a large Lorentz boost due to the heavy neutral Higgs boson. This feature leads to three collinear jets (a fat jet) which is a discriminating tool for identification of the top quarks from the Higgs boson resonances. Events with two identified top jets are selected and the invariant mass of the top pair is calculated for both signal and background. It is shown that the low $\tan\beta$ region has still some parts which can be covered by this analysis and has not been excluded yet by flavour physics data. 
\end{abstract}

\section{Introduction}
\label{Introduction}
The standard model (SM) of particle physics has taken a major step forward by observing the Higgs boson at LHC \cite{HiggsObservationCMS,HiggsObservationATLAS} based on a theoretical framework known as the Higgs mechanism  \cite{Englert1,Higgs1,Higgs2,Kibble1,Higgs3,Kibble2}. The observed particle may belong to a single SU(2) doublet (SM) or a two Higgs doublet model (2HDM)  \cite{2hdm1,2hdm2,2hdm3} whose lightest Higgs boson respects the observed particle properties.

One of the motivations for the two Higgs doublet model is supersymmetry where each particle has a super-partner. The supersymmetry provides an elegant solution to the gauge coupling unification, dark matter candidate and the Higgs boson mass radiative correction by a natural parameters tuning. In such a model two Higgs doublets are required to give mass to the double space of the particles \cite{MSSM1,MSSM2,MSSM3}. 

There are four types of 2HDMs with different scenarios of Higgs-fermion couplings. The ratio of vacuum expectation values of the two Higgs doublets ($\tan\beta=v_2/v_1$) is a measure of the Higgs-fermion coupling in all 2HDM types \cite{tanbsignificance}.

In general, 2HDM involves five physical Higgs bosons due to the extended degrees of freedom added to the model by introducing the second Higgs doublet. The lightest Higgs boson, $h$, is like the SM Higgs boson. The rest are two neutral Higgs bosons, $H,~A$ (subjects of this study), and two charged bosons, $H^{\pm}$. A review of the theory and phenomenology of 2HDM can be found in \cite{2hdm_TheoryPheno}. 

In addition to direct searches for the 2HDM Higgs bosons at colliders, there are indirect searches based on flavor Physics data by investigating sources of deviations from SM when processes which involve 2HDM Higgs bosons are introduced \cite{FMahmoudi}. Limites obtained from these type of studies are one of the strongest limits on the mass of the charged and neutral Higgs bosons and $\tan\beta$ and will be referred to when presenting the final results.

The adopted scenario in this analysis is a search for heavy neutral Higgs boson with mass in the range 0.5-1 TeV at LHC operating at $\sqrt{s}=14$ TeV. All heavy Higgs bosons (CP-even, CP-odd and the charged Higgs) are assumed to be degenerate, i.e., $m_H=m_A=m_{H^{\pm}}$. The region of interest is low $\tan\beta$ and the final restuls will be limited to $\tan\beta<2$. The signal process is $pp \to H/A \to t\bar{t} \to W^+bW^-\bar{b} \to jjbjjb$. The fully hadronic final state is expected to result in two fat jets (each consisting of three sub-jets associated with the top quark) which are examined using the updated \texttt{HEPTopTagger 2} \cite{HEPTopTagger1,HEPTopTagger2}. Events which contain two identified (tagged) top jets are used to fill the top pair invariant mass distribution histogram. The same approach is applied on background events and a final shape discrimination is performed to evaluate the signal significance. Before going to the details of the analysis, a brief review of the theoretical framework is presented in the next section.

\section{The Higgs sector of 2HDM}
The 2HDM Lagrangian for neutral Higgs-fermion couplings as introduced in \cite{2hdm_HiggsSector1} takes the form:
%\begin{widetext}
\begin{align}
\begin{split}
\mathcal{L}_Y&=\frac{1}{\sqrt{2}}\sum_{f} \bar{f}\left[\kappa^f s_{\beta-\alpha}+\rho^f c_{\beta-\alpha}\right]fh\\
&+\frac{1}{\sqrt{2}}\sum_{f} \bar{f}\left[\kappa^f c_{\beta-\alpha}- \rho^f s_{\beta-\alpha}\right]fH\\
&+\frac{i}{\sqrt{2}}\bar{f}\gamma_5\rho^f fA \\
\end{split}
\label{lag}
\end{align}
%\end{widetext}
with $U(D)$ being the up(down)-type quarks, $L$ the lepton fields, $h,~H,~A$ the neutral Higgs boson fields, $\kappa^f=\sqrt{2}\frac{m_f}{v}$ for any fermion type $f$ and $s_{\beta-\alpha}=\sin(\beta-\alpha)$ and $c_{\beta-\alpha}=\cos(\beta-\alpha)$. The $\rho^f$ parameters define the model type and are proportional to $\kappa^f$ as in Tab. \ref{types} \cite{Barger_2hdmTypes}. Therefore the four types of interactions (2HDM types) depend on the values of $\rho^f$ \cite{2hdm_HiggsSector2}. 

In this study, we require $s_{\beta-\alpha}=1$ which has two advantages. The first one is that the $s_{\beta-\alpha}$ factor in the lightest Higgs-gauge coupling is set to unity while the heavier Higgs, $H$, decouples from gauge bosons \cite{2hdm_TheoryPheno}. On the other hand, the SM-like Higgs-fermion interactions are $\tanb$ independent.
\begin{table}
\centering
\begin{tabular}{|c|c|c|c|c|}
\hline
\multicolumn{5}{|c|}{Type}\\
& I & II& III&IV\\
\hline
$\rho^D$ & $\kappa^D \cot\beta$ &$-\kappa^D \tan\beta$ &$-\kappa^D \tan\beta$ &$\kappa^D \cot\beta$  \\
\hline
$\rho^U$ & $\kappa^U \cot\beta$ &$\kappa^U \cot\beta$ &$\kappa^U \cot\beta$ &$\kappa^U \cot\beta$  \\
\hline
$\rho^L$ & $\kappa^L \cot\beta$ &$-\kappa^L \tan\beta$ &$\kappa^L \cot\beta$ &$-\kappa^L \tan\beta$  \\
\hline
\end{tabular}
\caption{Different types of 2HDM in terms of the Higgs boson couplings with $U$(up-type quarks), $D$(down-type quarks) and $L$(leptons).\label{types}}
\end{table}

\begin{figure}
\centering  \includegraphics[width=0.45\textwidth]{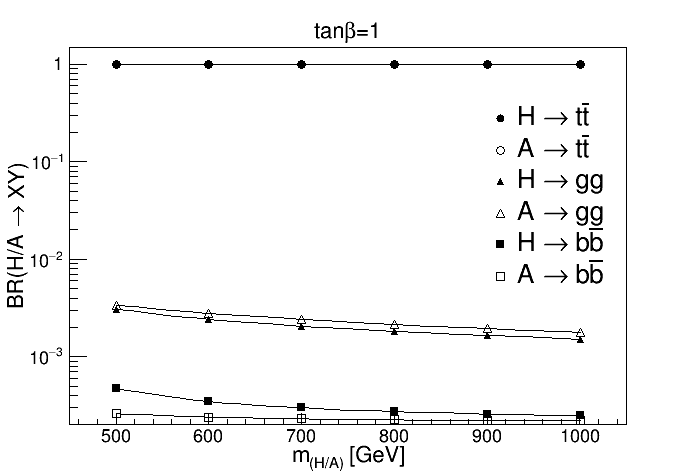}
  \caption{The branching ratio of neutral Higgs boson decays as a function of the mass. The $\tan\beta$ is set to 1.}
  \label{BRH_vs_mass}
\end{figure}

\begin{figure}
\centering  \includegraphics[width=0.45\textwidth]{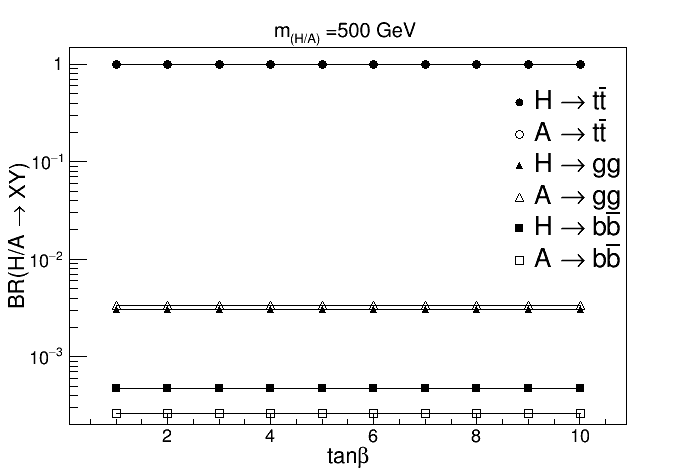}
  \caption{The branching ratio of neutral Higgs boson decays as a function of $\tan\beta$. The Higgs boson mass is set to 500 GeV.}
  \label{BRH_vs_tb}
\end{figure}

According to Tab. \ref{types}, the type I is interesting for low $\tanb$ as all couplings in the neutral Higgs sector are proportional to $\cot\beta$. This feature leads to cancellation of this factor as long as Higgs boson branching ratio of decay to leptons and quarks is concerned. The mass of the fermion thus plays an important role in the decay rate and as seen from Figs. \ref{BRH_vs_mass} and \ref{BRH_vs_tb}, the Higgs boson decay to $t\bar{t}$ dominates for all relevant Higgs boson masses and $\tan\beta$ values. The decay to a pair of gluons proceeds through a preferably top quark loop and stands as the second channel. The third channel is $H/A \to b\bar{b}$ which has been shown to be visible at LHC \cite{TypeI_LHC}. The current study focuses on $H/A \to t\bar{t}$ with branching ratio being near unity and independent of the Higgs boson mass (Fig. \ref{BRH_vs_mass}) and $\tan\beta$ (Fig. \ref{BRH_vs_tb}).

\section{Signal and background cross sections}
The signal process under study is a Higgs boson production with the Higgs boson masses in the range $500-1000$ GeV. The three Higgs bosons masses are set to be equal for minimizing $\Delta\rho$ \cite{drho}. All selected points are checked to be consistent with the potential stability, perturbativity and unitarity requirements and the current experimental limits on Higgs boson masses using $\texttt{2HDMC 1.6.3}$ \cite{2hdmc1,2hdmc2}.

There has been phenomenological searches for leptophilic Higgs boson within type IV 2HDM at LHC \cite{L2HDMLHC} and linear colliders \cite{hashemi2017,hashemi2017_2}. These searches are based on leptonic decay of the Higgs boson. On the other hand, the type I 2HDM can be considered as a leptophobic model where the Higgs boson decay to quarks plays an important role. At the first glance, decays to all fermions are relevant at low $\tan\beta$ values. However, the fermion mass in the Higgs-fermion vertex enhances the top quark coupling dramatically compared to other channels. This is due to the fact that the common $\cot\beta$ factors cancel out when calculating branching ratio of Higgs decays to fermions. Therefore in this analysis, the Higgs boson decay to $t\bar{t}$ is considered as the signal. 

While the neutral Higgs boson searches at LEP \cite{lep1,lep2} leads to $m_{A}\geq93.4$ GeV, the LHC results \cite{CMSNeutralHiggs,ATLASNeutralHiggs} indicate that the neutral Higgs boson mass in the range $m_{H/A}=200-400$ GeV is excluded for $\tan\beta \geq 5$. This result is based on minimal supersymmetric standard model (MSSM) which has a different Higgs boson spectrum from 2HDM due to supersymmetry constraints. Since our region of interest is Higgs boson masses above 500 GeV, no contraints from LEP or LHC limits the current analysis and the Higgs boson masses under study.
 
There are also results from flavor physics data which impose lower limits on the charged Higgs mass in type II and III at 480 GeV \cite{Misiak}. An update to this work is reported in \cite{misiak2017} where low $\tan\beta$ values are excluded to some extent. The idea in such analyses is based on the contribution from additional Feynman diagrams which involve charged Higgs bosons and their effect depends on the type of the 2HDM. The type I and IV behave different from type II and III as far as the charged Higgs coupling to quarks is concerned. In the former, the charged Higgs coupling to all quark types is suppressed at low $\tan\beta$, while in the latter, coupling with at least one type of the quarks (up type or down type) is enhanced with $\tan\beta$. Therefore charged Higgs limits from flavor physics in type I and IV are very soft and basically relevant at $\tan\beta$ values as low as 2. This is the region of search in this analysis. Although we are dealing with neutral Higgs bosons, since the scenario under study is a degenerate scenario based on $m_H=m_A=m_{H^{\pm}}$, limits on the charged Higgs are propagated into the final results. 

\begin{figure}
\centering
  \includegraphics[width=0.45\textwidth]{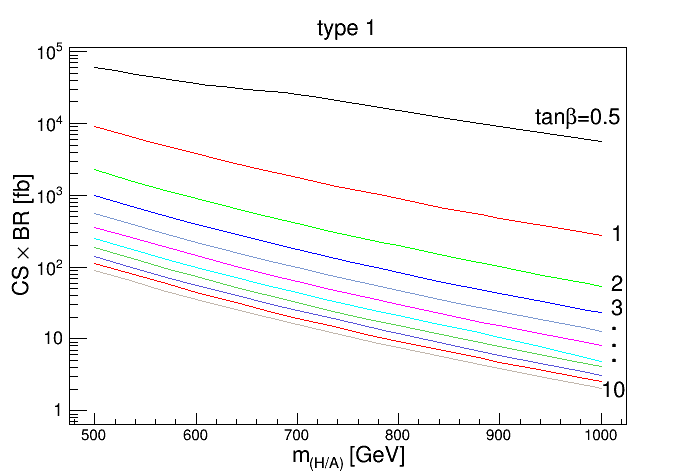}
  \caption{The signal cross section times BR$(H/A\to t\bar{t})$ at $\sqrt{s}=14$ TeV as a function of the Higgs boson mass.}
  \label{CS_vs_mass}
\end{figure}

\begin{figure}
\centering
  \includegraphics[width=0.45\textwidth]{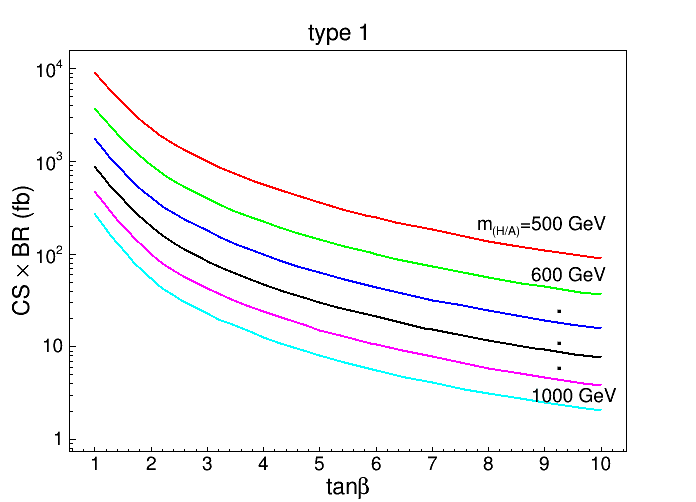}
  \caption{The signal cross section times BR$(H/A\to t\bar{t})$ at $\sqrt{s}=14$ TeV as a function of $\tan\beta$.}
  \label{CS_vs_tb}
\end{figure}

The signal cross sections times branching ratio of Higgs ($H/A$) decay to $t\bar{t}$ are shown in figs. \ref{CS_vs_mass} and \ref{CS_vs_tb}. The cross section decreases with increasing the Higgs boson mass as well as $\tan\beta$. Therefore the most suitable area for search is where the mass is as low as possible and $\tan\beta$ is also very small. 

The main SM background processes are $t\bar{t}$, gauge boson pair production $WW$, $WZ$, $ZZ$, $s-$channel and $t-$channel single top, single $W$ and single $Z/\gamma^*$. The signal and background cross sections are listed in tab. \ref{cs}.
\begin{table*}[h]
\centering
\begin{tabular}{|c|c|c|c|c|c|c|}
\hline
& \multicolumn{6}{c|}{Signal}\\
\hline
$m_{(H/A)}$ [GeV] & 500 & 600 & 700 & 800 & 900 & 1000 \\
\hline
$\sigma \times BR$ [fb] & 61.1 & 36.0 & 25.4 & 15.0 & 9.1 & 5.6 \\
\hline
\end{tabular}
\centering
\begin{tabular}{|c|c|c|c|c|c|c|c|c|}
\hline
& \multicolumn{8}{c|}{Background}\\
\hline
 & tt & WW & WZ & ZZ & sts & stt & W & Z \\
\hline
$\sigma \times BR$ [pb] & 390 & 32.6 & 12.1 & 5.33 & 5.6 & 117 & 1.02$\times 10^5$ & 5.7$\times 10^4$\\
\hline
\end{tabular}
\caption{The signal and background cross sections at $\sqrt{s}=14$ TeV. The ``sts'' and ``stt'' denote the $s$-channel and $t$-channel single top processes respectively.}
\label{cs}
\end{table*}
\section{Signal selection and analysis}
The generation of signal and background events starts with {\tt PYTHIA 8} \cite{pythia} followed by jet reconstruction using {\tt FASTJET 2.8} \cite{fastjet1,fastjet2}. 

The jet reconstruction algorithms are classified according to their different subjet distance measures which can be written as $d_{j1j2}=\Delta R^2_{j1j2}/D2 \times \textnormal{min}(p_{T,j1}^{2n},p_{T,j2}^{2n})$ with $n=-1,0,1$ for anti-$k_T$, Cambridge/Aachen (CA) and $k_T$ algorithms respectively. The $k_{T}$ algorithm first combines the soft and collinear subjets and is suitable for reconstructing the QCD splitting history in top tagging algorithm. The anti-$k_T$ algorithm, first combines the hardest subjets to obtain a stable jet with clean jet boundary. The CA algorithm always combines the most collinear subjets while not being sensitive to soft splittings and therefore is suitable for top tagging reconstruction. The algorithm adopted by \texttt{HEPTopTagger} is thus CA with a cone size of $\Delta R=1.5$.

The \texttt{HEPTopTagger} is one of recent algorithms introduced for boosted top quark reconstruction \cite{tt1}. It is based on a CA jet reconstruction with $\Delta R=1.5$ and the top jet candidate $p_T$ above 200 GeV. The threshold can be lowered down to 150 GeV without significant loss of efficiency \cite{jss8,Hashemi:2015hxc}. Having the collection of fat jets in the first step, the top tagging algorithm starts with undoing the last clustering of the top jet candidate $j$ and requiring the mass drop criterion as min$m_{j_{i}}<0.8 m_j$ where $j_i$ is the $i$th subjet from the jet $j$.  Subjets with $m_{j}<30$ GeV are not considered to end the unclustering iteration. 

In the second step a filtering is applied to find a three-subjet combination with a jet mass within $m_t\pm25$ GeV. 

In the last step, having sorted jets in $p_T$, several requirements are applied to find the best combination of subjets with two subjets giving the best $W$ boson invariant mass and the whole three subjets to be consistent with the top quark invariant mass. Details of these criteria are expressed in \cite{jss8}. 

\begin{figure}
\centering  \includegraphics[width=0.45\textwidth]{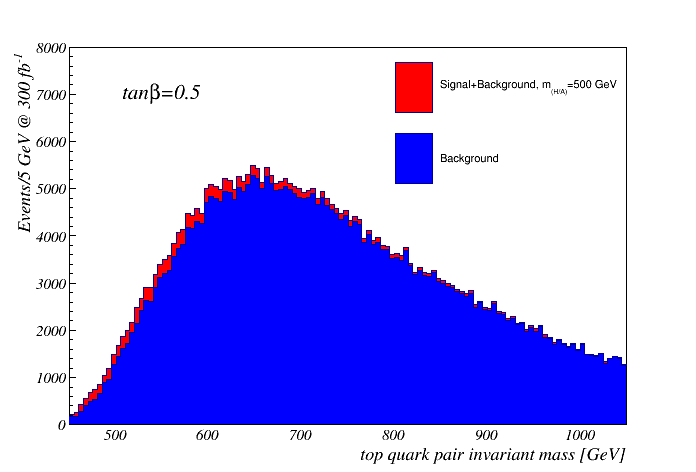}
  \caption{Signal (red) on top of the standard model background (blue) with $m_{(H/A)}=500$ GeV.}
  \label{m500}
\end{figure}

\begin{figure}
\centering  \includegraphics[width=0.45\textwidth]{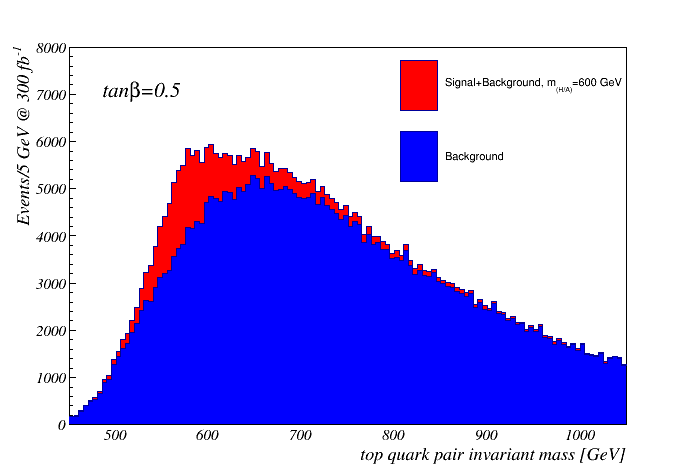}
  \caption{Signal (red) on top of the standard model background (blue) with $m_{(H/A)}=600$ GeV.}
  \label{m600}
\end{figure}

\begin{figure}
\centering  \includegraphics[width=0.45\textwidth]{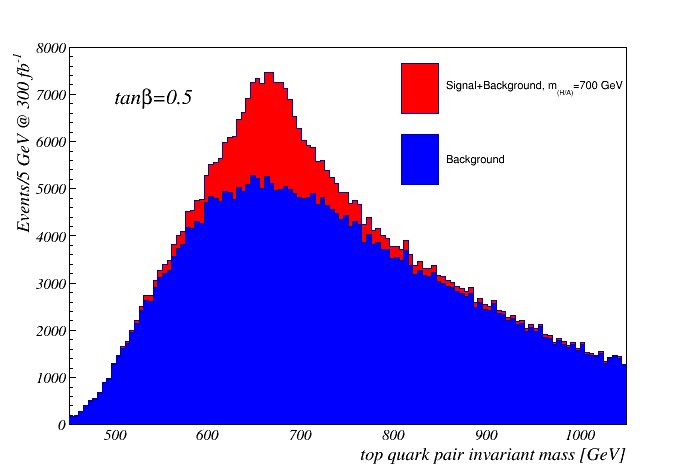}
  \caption{Signal (red) on top of the standard model background (blue) with $m_{(H/A)}=700$ GeV.}
  \label{m700}
\end{figure}

\begin{figure}
\centering  \includegraphics[width=0.45\textwidth]{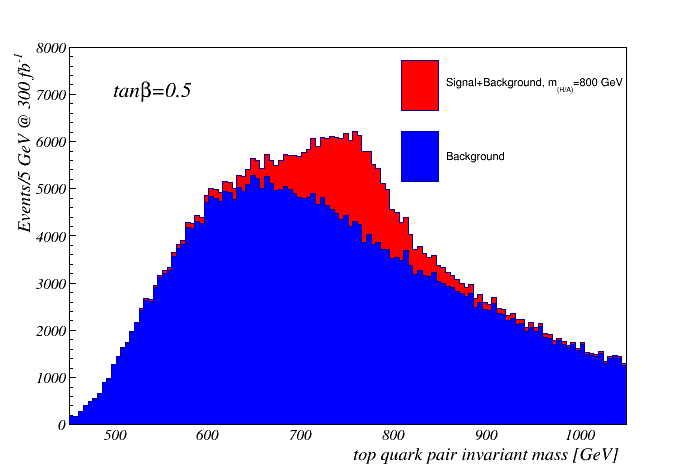}
  \caption{Signal (red) on top of the standard model background (blue) with $m_{(H/A)}=800$ GeV.}
  \label{m800}
\end{figure}

\begin{figure}
\centering  \includegraphics[width=0.45\textwidth]{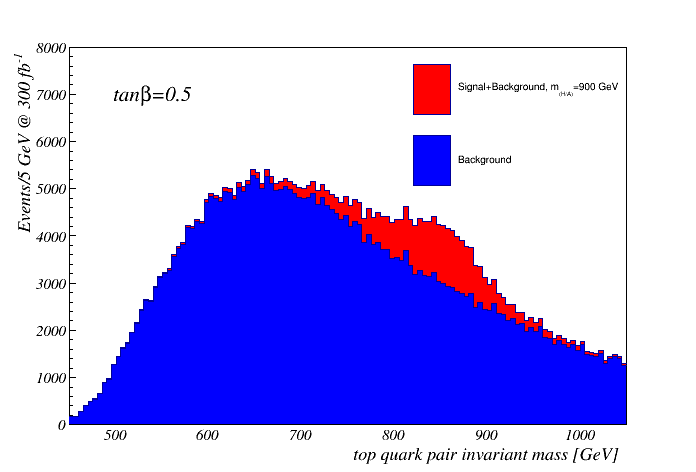}
  \caption{Signal (red) on top of the standard model background (blue) with $m_{(H/A)}=900$ GeV.}
  \label{m900}
\end{figure}

\begin{figure}
\centering  \includegraphics[width=0.45\textwidth]{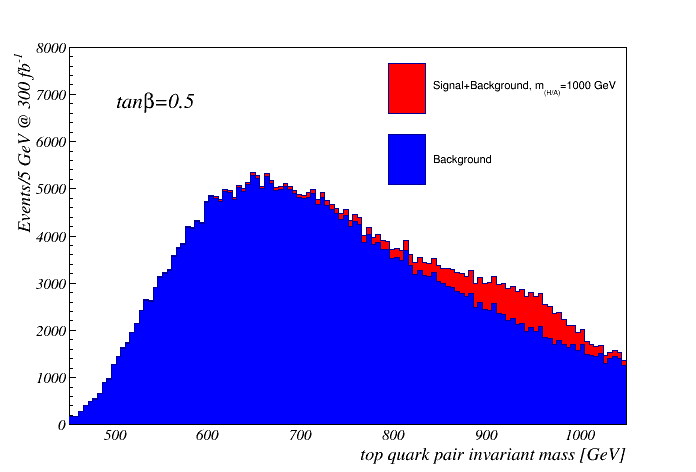}
  \caption{Signal (red) on top of the standard model background (blue) with $m_{(H/A)}=1000$ GeV.}
  \label{m1000}
\end{figure}

Performing the algorithm, a selection efficiency for each signal sample is obtained. The same procedure is applied on background samples. An event is required to have two top jets identified. The invariant mass of the two top jets are calculated as the Higgs boson candidate mass. Both signal and background distributions of top quark pair invariant masses are normalized according to the corresponding cross sections. The signal on top of the background is then plotted for each benchmark point as seen in Figs. \ref{m500}-\ref{m1000}.

At this step, since a large number of background is still filling the signal region, a mass window is applied to select the signal and increase the signal to background ratio. The position of the mass window (both left and right sides) is determined in an automatic search based on requiring the maximum signal significance. This is performed in a loop over bins of the histogram and finding the left and right bins inside which the signal significance is maximum. 

Table \ref{effs} shows mass window position, total efficiencies for signal and background events, final number of signal and background events passed the mass window cut, their ratio and the signal significance as $S/\sqrt{B}$ at two values of $\tan\beta=0.5$ and 1. The integrated luminosity is set to 300 $fb^{-1}$. The tab. \ref{effs} clearly shows the high sensitivity of the signal significance to $\tan\beta$ parameter. The analysis is thus relevant to $\tan\beta$ values as low as $\sim 2$. 

\begin{figure}
\centering
  \includegraphics[width=0.45\textwidth]{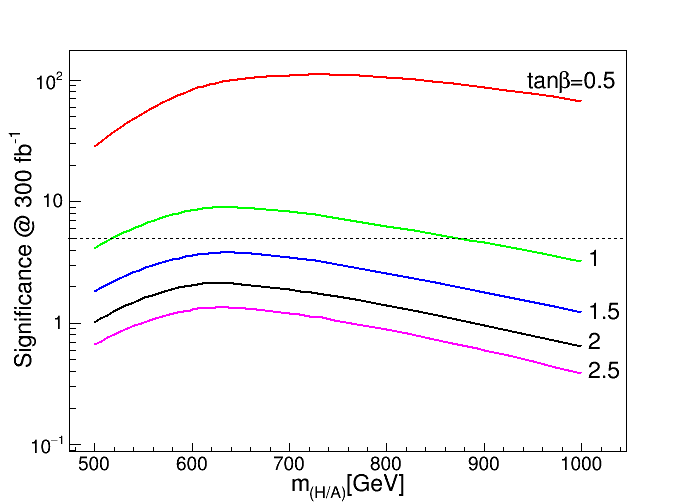}
  \caption{The signal significance at 300 $fb^{-1}$ as a function of the Higgs boson mass for different vlaues of $\tan\beta$}
  \label{Significance}
\end{figure}

\begin{figure}
\centering
  \includegraphics[width=0.45\textwidth]{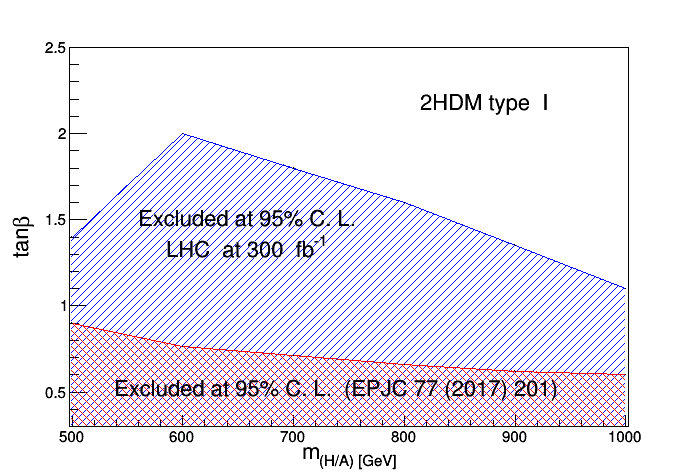}
  \caption{The $95\%$ C.L. exclusion region at 300 $fb^{-1}$.}
  \label{2sigma}
\end{figure}

\begin{figure}
\centering
  \includegraphics[width=0.45\textwidth]{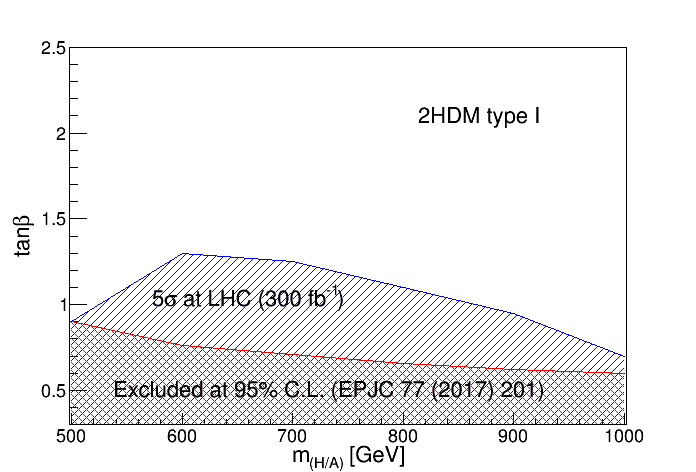}
  \caption{The $5\sigma$ discovery region at 300 $fb^{-1}$.}
  \label{5sigma}
\end{figure}

Figure \ref{Significance} shows the signal significance as a function of the Higgs boson mass for different $\tan\beta$ values. The dashed horizontal line indicates the $5\sigma$ significance. Using the analysis results for Higgs boson masses from 500 GeV to 1000 GeV, one can obtain the 95$\%$ C.L. exclusion region and the $5\sigma$ discovery contours. Figure \ref{2sigma} shows the exclusion region at $95\%$ C.L. including the recent result from \cite{misiak2017} (the result reported in \cite{misiak2017} is based on charged Higgs mass as a function of $\tan\beta$, however, it is included in the current work as a limit for all Higgs bosons since the Higgs boson masses are equal in the scenario adopted in this analysis). The $5\sigma$ contour is also shown in Fig. \ref{5sigma}. 

As seen from Figs. \ref{2sigma} and \ref{5sigma}, both exclusion and discovery are possible at regions not yet excluded by any experimental or phenomenological analysis. Therefore any sign of extra top pair signals on top of SM background could be regarded as a signal for new physics especially 2HDM. It should be noted that in this analysis, a full set of background processes was studied. However, all background processes led to very small number of events which were negligible compared to the SM $t\bar{t}$. Therefore final plots are based on signal on top of the $t\bar{t}$ distribution without any sizable error. 

\section{Conclusions}
Extra sources of $t\bar{t}$ events from what we expect from standard model can appear from theories beyond standard model such as two Higgs doublet models. In 2HDM type I, the heavy neutral (CP-even or odd) Higgs decay to $t\bar{t}$ dominates the other channels. In such a scenario a proton-proton collision may create a neutral Higgs decaying to $t\bar{t}$. The signal from such a process, can be observed as an excess of top pair events over what is expected from SM. The discriminating tool can be a top pair invariant mass distribution filled with events containing two top jets from both signal and background processes. The analysis performed in this work, shows that such a signal is observable at integrated luminosity of 300 $fb^{-1}$ for $\tan\beta$ values which depend on the Higgs boson mass. The exclusion at 95$\%$ C.L. is also possible at the same integrated luminosity for $\tan\beta<2$ with $m_{(H/A)}=600$ GeV as the best point.
 
\def\reversed#1&#2&#3&#4&#5&#6&#7\\{#7&#6&#5&#4&#3&#2&#1\\}
 \begin{table*}[h!]
 	\centering
 	\caption{Signal and background analysis results at $\tan\beta=0.5$ and 1}
 \begin{tabular}{|c|c|c|c|c|c|c|}
  \hline
     & \multicolumn{6}{c|}{$m_{(H/A)}$ [GeV] }\\
        \hline
\reversed    $1000$&$900$&$800$&$700$&$600$&$500$&\multicolumn{1}{|c|}{}  \\
     \hline  
\reversed $865-1020$ & $780-925$ &  $695-825$ & $610-730$  &  $525-630$ & $450-635$  &  Mass window [GeV]\\
\hline
\reversed $0.016$ & $0.014$ &  $0.012$ & $0.008$  &  $0.0036$ & $0.0009$  &  Tot-eff(S)\\
\hline
\reversed $0.0037$ & $0.0037$ &  $0.0037$ & $0.0037$  &  $0.0037$ & $0.0037$  &  Tot-eff(B)\\
\hline
\reversed $17315$ & $26295$ &  $35225$ & $38365$  &  $24092$ & $9052$  &  $S$\\
\hline
\reversed $67870$ & $91049$ &  $112781$ & $123454$  &  $84076$ & $101998$  &  $B$\\
\hline
\reversed $0.255$ & $0.29$ &  $0.075$ & $0.312$  &  $0.287$ & $0.0887$  &  $\frac{S}{B}$\\
\hline
\reversed \multirow{2}{*}{66.462}  & \multirow{2}{*}{87.14} &  \multirow{2}{*}{104.88} & \multirow{2}{*}{109.2}  &  \multirow{2}{*}{83.9} & \multirow{2}{*}{28.34} &  $\frac{S}{\sqrt{B}}$\\
$\tan\beta=0.5$ & & & & & & \\
\hline
\reversed \multirow{2}{*}{3.18}  & \multirow{2}{*}{4.57} &  \multirow{2}{*}{6.25} & \multirow{2}{*}{8.24}  &  \multirow{2}{*}{8.49} & \multirow{2}{*}{4.14} &  $\frac{S}{\sqrt{B}}$\\
$\tan\beta=1$ & & & & & & \\
\hline
 \end{tabular}\label{effs}
 \end{table*}


\section*{Acknowledgements}
This work was performed using the computing cluster at Shiraz University, college of sciences. We would like to thank the personnel involved in the operation and maintenance of the cluster.\\

\bibliographystyle{spphys}
\bibliography{BIB_TO_USE}

\end{document}